# Hydrogen response to high-density dislocations in bulk perovskite oxide SrTiO$_3$

Xufei Fang[1*,] Lars Dörrer[2], Svetlana Korneychuk[1,3], Maria Vrellou[1], Alexander Welle[3,4], Stefan Wagner[1], Astrid Pundt[1], Harald Schmidt[2*], Christoph Kirchlechner[1]

[1]Institute for Applied Materials, Karlsruhe Institute of Technology, 76131 Karlsruhe, Germany

[2]Institute of Metallurgy/Solid-State-Kinetics Group, Clausthal University of Technology, 38678 Clausthal-Zellerfeld, Germany

[3]KNMFi, Karlsruhe Institute of Technology, 76131 Karlsruhe, Germany

[4]Institute of Functional Interfaces, Karlsruhe Institute of Technology, 76344 Eggenstein-Leopoldshafen, Germany

*Corresponding authors: xufei.fang@kit.edu (X.F.); harald.schmidt@tu-clausthal.de (H.S.)

**Abstract**

Hydrogen plays an increasingly important role in green energy technologies. For instance, proton-conducting oxides with high performance for fuel cell components or electrolysers need to be developed. However, this requires a fundamental understanding of hydrogen-defects interactions. While point defects and grain boundaries in oxides have been extensively studied, the role of dislocations as line defects remains less understood, primarily due to the challenge for effective dislocation engineering in brittle oxides. In this work, we demonstrate the impact of dislocations in bulk single-crystal perovskite oxide SrTiO$_3$ on hydrogen uptake and diffusion using deuterium as tracer. Dislocations with a high density up to ~$10^{14}$/m$^2$ were mechanically introduced at room temperature. Exposing this dislocation-rich and the reference regions (with a dislocation density of ~$10^{10}$/m$^2$) to deuterium at 400 °C for 1h, followed by secondary ion mass spectrometry measurements, we observed a ~100 times increase in deuterium incorporation in the dislocation-rich region. The result suggests that dislocations in oxides can act as an effective reservoir for deuterium. This proof-of-concept brings new insights into the emerging hydrogen-dislocation interactions in functional oxides.

**Graphical abstract**

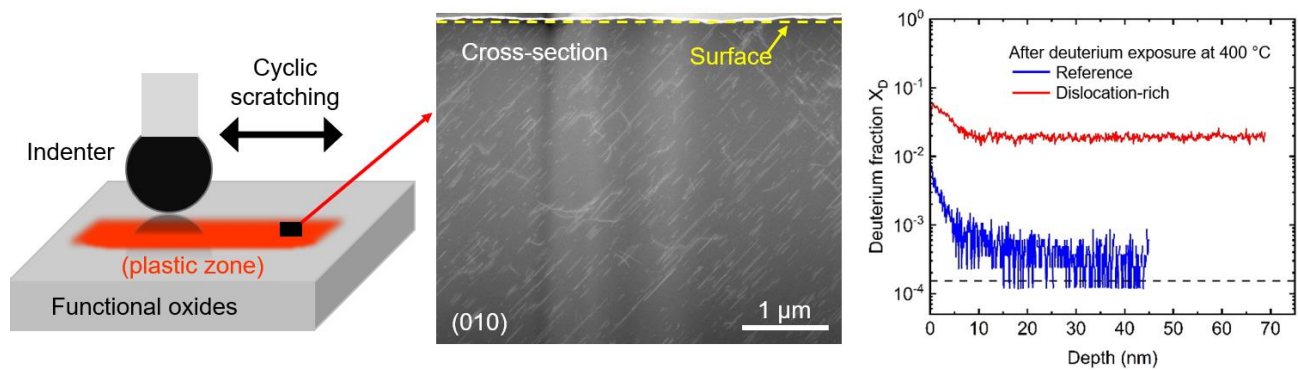

## 1. Introduction

Hydrogen plays an increasingly important role due to its high societal impact in green energy technologies, such as hydrogen production and storage for energy conversion, green steels manufacturing for $CO_2$ reduction [1], and energy efficient computing [2]. Parallel to the extensive efforts in understanding hydrogen embrittlement in metallic materials for hydrogen storage and transportation, developing proton-conducting oxides with high performance for electrolysers (for hydrogen production) and fuel cells (for hydrogen consumption) is sharing a fair fraction of scientific and economic weight. This requires fundamental understandings of hydrogen-defects interactions in oxides. So far, the majority of defect engineering efforts have been focused on point defects (e.g., oxygen vacancy tuning [3]) or grain boundaries as 2D defects (e.g. promoting the acceptor dopants segregation [4]). Another major defect type, dislocations as line defects, in proton-conducting oxides or for hydrogen production has rarely been considered, primarily due to the perception that ceramics are brittle and do not exhibit dislocation activities. This conventional picture is being revisited in recent years, as an increasing number of ceramics are being demonstrated for dislocation-mediated plastic deformation at room temperature up to meso/macroscale [5], as well as the dislocation-tuned functional and mechanical properties [6-9].

Dislocations as line defects can not only carry plastic deformation but also may act as fast diffusion paths for hydrogen [10]. In the meantime, novel proofs-of-concepts have been put forward showcasing the enhanced hydrogen production rate (via water splitting) in functional oxides such as $TiO_2$ nanoparticles [11] and bulk $BaTiO_3$ [12] with pre-engineered dislocations. Enhancement in proton conductivity due to misfit dislocations has also been reported via strain engineering in thin oxide films [13, 14]. Nevertheless, these studies require delicate processing procedures at the expense of high energy cost, e.g., high-temperature bulk compression at 1150 °C with an intermediate dislocation density of ~$10^{12}/m^2$ [12], or have limited degree of controlling the dislocations in nanoparticles or at the interfaces of thin films. In terms of hydrogen-dislocation interaction (for instance, segregation, trapping, and diffusion) in functional oxides, the basic understandings remain far less developed [10, 15, 16]. To this end, the first challenge is how to effectively and efficiently engineer high-density dislocations, preferably with much higher density than previously achieved [12] in a large volume populated with dislocations.

In this study, we aim to address the following questions: How does hydrogen respond to high-density dislocations in a bulk oxide provided that the dislocations can be successfully engineered? Are these dislocations stable at elevated temperatures for potential operation scenarios in e.g. solid oxide cells?  In what follows, we adopted single-crystal $SrTiO_3$ as a prototypical perovskite oxide, which exhibits room-temperature dislocation plasticity in bulk [17] and allows for efficient dislocation imprinting with high densities up to ~$10^{14}/m^2$ at room temperature [8, 18]. Then the sample, containing dislocation-rich as well as reference undeformed region, was exposed to a hydrogen isotope, deuterium-rich environment at elevated temperatures, followed by secondary ion mass spectrometry (SIMS) measurements to assess the deuterium diffusion profile.

## 2. Results and Analyses

### *2.1. Dislocation engineering and visualization*

A nominally undoped SrTiO$_3$ single crystal with (001) surface orientation was used (Alineason GmbH, Frankfurt am Main, Germany). Starting with a sample dimension of 5 mm × 5 mm × 1 mm, we mechanically imprinted dislocations in an area of about 2.5 × 2.5 mm$^2$ (**Fig. 1a**), which is suitable for SIMS measurement. The dislocations were introduced by cyclic scratching the sample surface at room temperature [19] with a Brinell indenter of a diameter of 2.5 mm and a normal load of 1 kg. The depth of the dislocations induced using this method extends beyond 50 μm [20]. Each individual wear track (**Fig. 1a,b**) has a maximum depth of ~300 nm over ~130 μm [19], yielding a nominally flat surface after scratching. Contrasting the reference region that has a dislocation density of ~10$^{10}$/m$^2$ [21], high-density dislocations (~10$^{14}$/m$^2$) were successfully generated in the scratch track after 10 passes of scratching, as evidenced by the TEM (transmission electron microscopy) observation of the cross section (**Fig. 1c**) along the scratch track (**Fig. 1b**). The dislocations are primarily aligned 45° with respect to the surface, corresponding to the activated {110} <110> slip systems in (001) SrTiO$_3$ at room temperature [18, 22]. Within the TEM field of view (deeper than 3 μm), the dislocations are distributed homogeneously along the depth in the near-surface region. No amorphization was detected in the heavily deformed region, which remains in single-crystal form as confirmed by the TEM diffraction pattern.

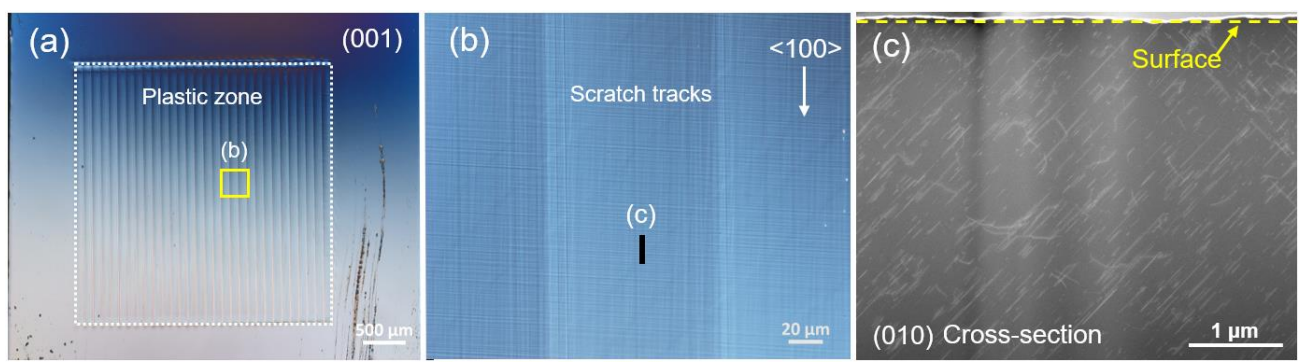

**Figure 1.** Visualization of the plastic zone and dislocations: (a) Overview of the large plastic zone generated by overlapping the scratch tracks on the (001) surface. (b) Zoomed-in view of three scratch tracks in parallel, corresponding to the yellow rectangle in (a). (c) TEM image of the cross-section region, lifted out at the surface, as indicted by the black line in (b).

## 2.2. Deuterium incorporation and SIMS measurements

We used deuterium (D) as a tracer for a better distinction from hydrogen (H) background. D uptake was achieved by annealing the samples in a furnace at 400 °C for 1 h in a wet $O_2/D_2O$ atmosphere (bubbler at room temperature). The furnace works under oxidizing conditions utilizing an oxygen flow of 6 L/h saturated with $D_2O$. $D_2O$ dissociates and D is incorporated into the material that presumably leads to the formation of hydroxyl (–OD) groups [23]. Oxygen vacancies may be involved in the incorporation process [24] and discussions will follow. The temperature of 400 °C was adopted to accelerate the D uptake without altering the dislocation structure, which will be confirmed later in the experiments. Hydrogen isotope (H and D) selective SIMS measurements (Cameca IMS-3F/4F) in depth profile mode were performed on the sample surface before and after the sample exposure to $D_2O$. An $O^-$ primary ion beam (14.5 keV, 30 - 90 nA) was used. The sputtered area was about 250 µm × 250 µm, wherefrom 20% in the centre was gated for further signal processing in a double focused mass spectrometer. In depth profiling mode, the secondary ion intensities of $H^+$ and $D^+$ ions were recorded as a function of sputtering time. Depth calibration was obtained by measuring the crater depth with a mechanical surface profiler (Tencor, Alphastep). To ensure a direct comparison, the tests were performed on one same sample containing both the dislocation-rich and dislocation-scarce regions (see **Fig. 1a**). Note that the isotopes H ($^1H$) and D ($^2H$) are chemically identical (neglecting the isotope effect), and the intensity of the respective SIMS signals ($I_H$ and $I_D$) is converted to calculate the relative D fraction to $X_D = I_D/(I_H+I_D)$.

As illustrated in **Fig. 2a,** prior to $D_2O$ exposure the relative deuterium fraction of the regions with a high dislocation density of ~$10^{14}/m^2$ (dislocation-rich) and a low dislocation density of ~$10^{10}/m^2$ (reference) is identical, overlapping with the natural isotope background of $X_D$ ~$1.5 \times 10^{-4}$ [25]. This serves as a benchmarking test. After $D_2O$ exposure (**Fig. 2b**), the relative deuterium fraction is enhanced by a factor of ~100 in the dislocation-rich region to a depth up to 70 nm, while in the region without dislocations it stays close to the natural background. This indicates that additional amount of deuterium is incorporated into the sample in the dislocation-rich region. The increase of the relative deuterium fraction at the sample surface for depths below ~10 nm indicates a modified surface structure (which is known for $ABO_3$-type perovskite oxides [26] even without treatment in hydrogen environment), where the deuterium may be more easily incorporated for both cases. This could be caused by the formation of a thin space charge layer at the surface [27] or near-surface reconstruction after annealing in the presence of deuterium, where a higher oxygen vacancy concentration is formed [28]. The conditions of SIMS measurements are the same for dislocation-free and dislocation-rich regions. However, as can be seen from **Fig. 2**, the probed depths for the reference and dislocation-rich regions are consistently different, before and after

D exposure. The difference can be explained by a higher sputter yield (material removal rate) due to the severely distorted lattices in dislocation-rich regions compared to the reference regions.

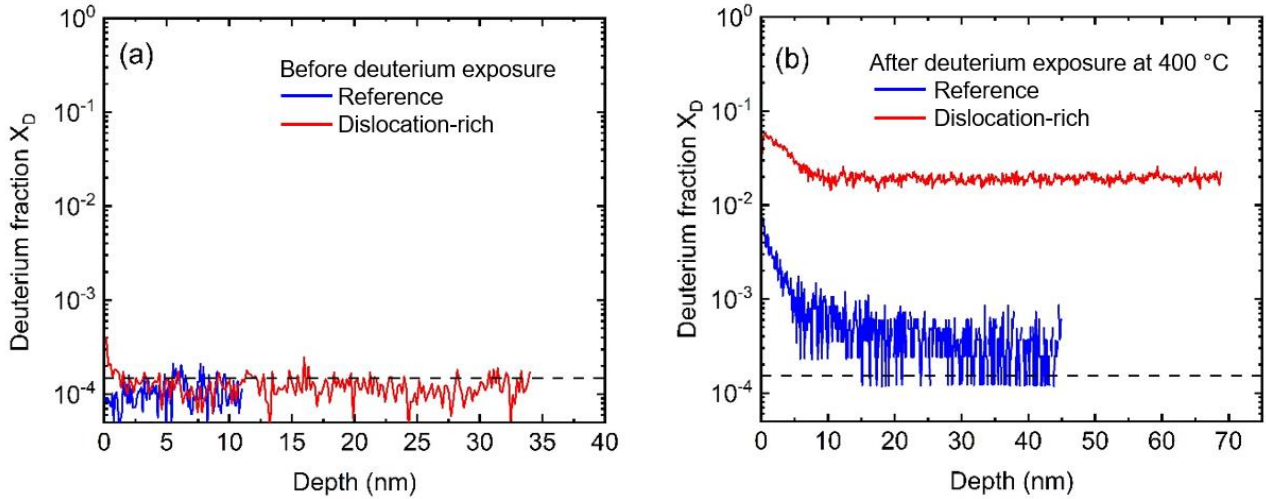

**Figure 2.** Deuterium fraction measured on a nominally undoped SrTiO$_3$ single crystal by SIMS in areas with a high dislocation density of ~10$^{14}$/m$^2$ (dislocation-rich) and a low dislocation density ~10$^{10}$/m$^2$ (reference): (a) before exposure to D$_2$O. (b) after exposure to D$_2$O at 400 °C for 1 h. The natural isotope background of ~1.5 × 10$^{-4}$ is indicated by the black dashed line.

### *2.3. Diffusivity of deuterium in the dislocation-rich region*

In order to gain insight into the modification of the deuterium fraction at larger depths and to determine the diffusivity of deuterium in the dislocation-rich region, long-time SIMS measurement was performed to depths of ~2700 nm using a higher primary beam intensity. This depth is within the TEM field of view which visualized a uniform dislocation density along the depth within ~3000 nm (**Fig. 1c**). As illustrated in **Fig. 3**, we observe a continuous penetration of deuterium into the sample in the dislocation-rich region and a decrease of the deuterium fraction over two orders of magnitude. In contrast, the reference region reveals a constant low level of D fraction up to 1000 nm.

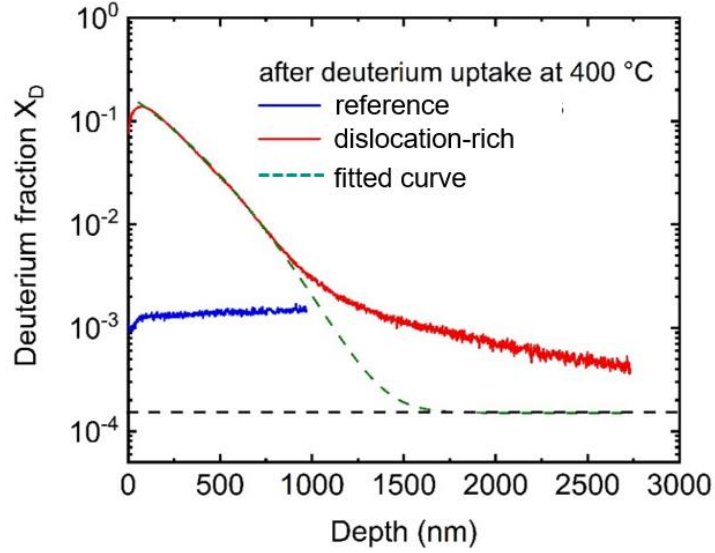

**Figure 3.** Deuterium fraction as a function of depth for larger depths. The fitting curve (green dashed line) according to Eq. (1) is presented. The natural isotope background is indicated by a black dashed line.

In the range between 70 and 900 nm (the near-surface data was cut off to avoid compositional uncertainties caused by surface roughness induced by the scratching), the diffusion profile can be described [29] by the following solution of Fick's second law using the boundary condition at the surface $-D\frac{\partial c}{\partial x} = K\,(c_0 - c_s), x = 0$ [30]:

$$c(x,t) - c_\infty = (c_0 - c_\infty)\left[\mathrm{erfc}\left(\frac{x}{\sigma}\right) - \exp\left(2\frac{x}{\sigma}\sqrt{\frac{t}{\tau}} + \frac{t}{\tau}\right)\mathrm{erfc}\left(\frac{x}{\sigma} + \sqrt{\frac{t}{\tau}}\right)\right] \quad (1)$$

with $c_s$ being the actual surface concentration, $\sigma = 2\sqrt{D_D t}$, and $\tau = D_D/K^2$, while $c_0 = 0.95$ is the relative fraction of D in the source, $c_\infty = 1.5 \times 10^{-4}$ is the natural abundance of D in the single crystal, $D_D$ is the deuterium diffusivity and $K$ is the surface-exchange coefficient. The quantity $K$ describes the kinetics of the exchange between the heavy water source and the crystal. We obtain from least-squares fitting values of $D_D = 2.7 \times 10^{-17}$ m²/s and $K = 1.7 \times 10^{-11}$ m/s. The relative error of the diffusivity is about 30% resulting from the determination of the crate depth. The deuterium diffusivity is rather low compared to other perovskite oxides like undeformed $LiNbO_3$, where such diffusivities were already reached at about 250 °C [31]. This is surprising because the present diffusion of hydrogen in $SrTiO_3$ along dislocations is expected to be fast. We speculate that trapping effects of other types of defects may play a considerable role, as will be discussed later.

The fitting curve in **Fig. 3** does not capture the long tail of the SIMS signal at depths higher than 900 nm, which hints to a second faster diffusion process responsible for transporting deuterium in low amounts deeper into the sample. Future investigations will include time and temperature dependent exposure to D$_2$O to address this point. Note that for the reference region of the sample (blue curve in **Fig. 3**), the deuterium fraction is enhanced by a factor of six over the natural background, in contrast to **Fig. 2b**, where it is a factor of about two. This is likely caused by the fact that the pumping time of the SIMS analysis chamber might influence the determination of the exact D fraction due to the presence of residual hydrogen or water and resulting interferences with H. Further, measurements at different regions on the same sample were done by SIMS after pumping significant time in high vacuum. Some of the near-surface stored D may have diffused out prior the analysis performed later. The near-surface sites are always filled if they have lower energy: D is then diffusing from the sample's in-depth region into these near-surface sites. The lateral surface inhomogeneity in D uptake (see **Fig. 4b**) may also add to this fluctuation. Nevertheless, such fluctuations in the reference region are much less significant than the clear enhancement of D in the dislocation-rich region.

### *2.4. Retention of deuterium in the dislocation-rich region*

The 1 h incorporation of D at elevated temperature (400 ℃) as well as the low diffusivity obtained from **Fig. 3** suggests that once the D is incorporated in the dislocation-rich region, the majority of D can be trapped and remain in the dislocation-rich region for long time. To verify this, we performed additional time-of-flight SIMS (ToF-SIMS) measurements on the same sample after storing it under ambient conditions for about 300 days. The results in **Fig. 4** clearly illustrate D retention in the dislocation-rich region. **Figure. 4a** shows the scratch tracks (dislocation-rich regions) on the left half, and the reference regions on the right. The distribution of D fraction in **Fig. 4b** correlates excellently with **Fig. 4a**. Furthermore, a line profile (**Fig. 4c**) is extracted from **Fig. 4b** to display the D intensity profile as the distance from left to right as in **Fig. 4b**. Note the D intensity terminates almost abruptly at the end of the wear track, indicating that the dislocation distribution ends sharply at the end of the scratch track. This dislocation distribution was directly visualized by the Brinell indentation and dislocation etch pits analysis by Okafor et al. [32].

It is noted that $X_D = I_D/(I_H+I_D)$ calculated from depth integrated signal intensities after this long-time storage (~4.1 × 10$^{-3}$) is about one order of magnitude lower than the freshly exposed sample (**Fig. 2b**), suggesting a continuous but very sluggish diffusion of D out of the sample. In the untreated area $X_D$ was found to be ~4.5 × 10$^{-4}$, almost identical to the natural background, again confirming the impact of dislocations.

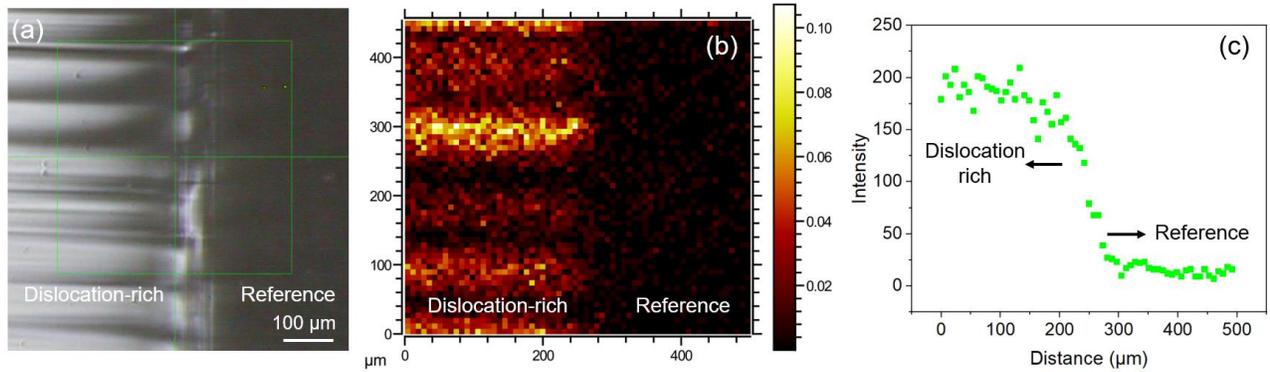

**Figure 4**. Direct correlation between the (a) optical image, (b) D fraction map, and (c) the quantified relative D intensity across the dislocation-rich and reference area. Note figure (b) corresponds to the large green rectangle indicated in (a).

## 3. Discussions

### 3.1. Increased hydrogen segregation at dislocations

Defects such as vacancies, dislocations, free surfaces, and grain boundaries can act as trapping sites for hydrogen, as widely established in crystalline solids with majority of the evidence in metallic bulk materials and thin films [33]. Hydrogen ad- and absorption at these defects can lower the system's total strain energy [10]. In the current case, the mechanically induced dislocations are effective for a much higher concentration of deuterium incorporation as confirmed by the SIMS measurements. It is expected that, the hydrogen uptake shall also be dependent on the dislocation densities, as will be investigated in detail in the future by tuning the dislocation densities in $SrTiO_3$ [34].

### 3.2. Thermal stability of the dislocations

As the D exposure was performed at 400 °C for 1 h, it raises the concern if the dislocations generated at room temperature have changed their configuration, namely, how thermally stable are these dislocations at 400 °C for 1 h. Previous thermal treatment of the dislocations in $SrTiO_3$ suggests that these line defects persist and do not annihilate out even up to 1200 °C [35, 36]. Negm et al. [35] showed evidence of dislocation motion and density increase subjected to annealing at 1100 °C for 1h. To further examine the thermal stability of dislocations, following the method by Negm et al. [35], we performed dislocation etch pit studies for samples treated at room temperature and up to 600 °C (higher than 400 °C in the current D exposure experiment) for 1h. The etch pits analysis for dislocations confirms no change of the dislocations structures in the surface region was observed at temperatures up to 600 °C (see **Fig. S1** in the Supplementary Materials). A comprehensive analysis of the temperature-dependent dislocation structure evolution merits an independent study and goes beyond the scope of this work.

*3.3. Impact of oxygen vacancies*

Though the focus of this work is hydrogen-dislocation interaction, it is pertinent to consider the impact of oxygen vacancies as the oxide samples were exposed to hydrogen-rich environment at elevated temperatures, and oxygen vacancies can be favorable trapping sites for hydrogen. Although commercially undoped $SrTiO_3$ is oxygen vacancy rich due to the tracer elements acting mostly as acceptor dopants [37], these oxygen vacancies in the reference sample are not relevant for the current hydrogen/deuterium detection as evidenced by the benchmarking SIMS measurement in **Fig. 2**.

However, with high-density dislocations mechanically generated in $SrTiO_3$, complexity is expected as previous reports suggest the dislocations in $SrTiO_3$ are easy to reduce [36], and the dislocations cores are oxygen deficient [38-41]. Rodenbücher et al. [42, 43] investigated the mechanically induced dislocations and their electrical conductivities, their results suggest a strong interaction between dislocations and oxygen vacancies formed due to reduction, with the latter being locally compensated by electrons. The increased concentration of oxygen vacancies bond to dislocations in $SrTiO_3$ may be responsible for the low diffusivity evaluated in **Sec. 2**, as these oxygen vacancies can act as effective trapping sites. The possible symbiosis of the dislocations and oxygen vacancies in $SrTiO_3$, together with the challenge in quantifying the oxygen vacancy concentrations in oxides [28, 44], makes it difficult at this stage to experimentally decouple these two types of defects to allow for *pure* dislocation-hydrogen interaction, which we expect that PALS (positron annihilation lifetime spectroscopy) [45] coupled with *ab initio* simulation may shed new light on it.

**4. Conclusion**

Mechanically engineered large area (~2.5 × 2.5 $mm^2$) with high-density dislocations (up to ~$10^{14}/m^2$) in nominally undoped single-crystal $SrTiO_3$ without forming cracks, sub-grain boundaries or amorphous phases allows the probing of the hydrogen response to dislocations via SIMS measurements, using deuterium as tracer. The high-density dislocations generated at room temperature remain thermally stable at least up to 600 °C for 1 h, and they host a much higher relative deuterium concentration by a factor of ~100 compared to the reference sample. A diffusivity of deuterium with ~2.7 × $10^{17}$ $m^2$/s has been obtained in the dislocation-rich region. Though this value is rather low, the findings can be relevant for hydrogen-production and proton-exchanging oxide ceramics if the dislocation mesostructure can be further tuned to enhance the diffusivity. Open questions remain particularly concerning the impact of different dislocation densities, dislocation structures, hydrogen induced defects [46], and their interaction with other defects, e.g., oxygen vacancies, dislocations, and interfaces on hydrogen uptake and diffusion.

**Acknowledgement**

X.F., M.V., and C.K. acknowledge the financial support by the European Research Council (ERC Consolidator Grant, project TRITIME, grant No. 101043969). X. F. is also funded by the ERC (ERC Starting Grant, project MECERDIS, grant No. 101076167) and the YIN Grant at Karlsruhe Institute of Technology. Views and opinions expressed are, however, those of the authors only and do not necessarily reflect those of the European Union or the European Research Council. Neither the European Union nor the granting authority can be held responsible for them. We thank A. Frisch for the optical image of the plastic zone and C. Kofahl for the annealing experiments in $D_2O$.

**Data availability:** The related data for this publication has been provided in the manuscript or in the Supplementary Materials.

**Conflict of Interest:** The authors declare no known conflict of interest.